\documentclass{aa}
\usepackage{graphicx}
\usepackage{amsmath}

\input{tcilatex}

\begin{document}
\thesaurus{12(12.07.1, 08.16.2)}

\title{Perturbative analysis in planetary\\
gravitational lensing}
\author{Valerio Bozza\thanks{%
E-mail valboz@physics.unisa.it}}
\institute{Dipartimento di Scienze Fisiche E.R. Caianiello,\\
 Universit\`{a} di Salerno, I-84081 Baronissi, Salerno, Italy.}
\date{Received / Accepted }
\maketitle

\begin{abstract}
The traditional perturbative method is applied to the case of gravitational
lensing of planetary systems. A complete and detailed description of the
structure of caustics for a system with an arbitrary number of planets can
be obtained. I have also found precise analytical expressions for
microlensing light curves perturbed by the presence of planets.
\end{abstract}

\section{Introduction}

\noindent In the last years, the prediction and the observation of many
microlensing events are gathering ever more interest in gravitational
lensing. The typical light amplification curve for these events, found by
Paczynski in 1986, has been observed by several
astronomical collaborations in observation campaigns toward the bulge of our
galaxy (Udalski et al., 1993a; Alard et al., 1997), the Large
Magellanic Cloud and the Small Magellanic Cloud (Alcock et al., 1993; Aubourg et al., 1993), the spiral arms and
Andromeda galaxy (Tomaney \& Crotts, 1996; Ansari et al., 1997; Melchior et al., 1999). Recently, observations toward
globular clusters have even been suggested (Jetzer, Strassle \& Wandeler, 1998). Beyond proving the
correctness of Paczynski's predictions, the observation of microlensing
events provides a very cunning instrument for the investigation of the halo
of our (and/or some other) galaxy.

Together with events strictly following Paczynski's curve, some events
showing deviations from the standard behaviour have been detected. Each of
these deviations has found some interpretation (Finite source cut-off
(Witt \& Mao, 1994; Alcock et al., 1998), blending (Sutherland, 1998), parallax effect
(Gould, 1992; Alcock et al., 1995), binary lens (Schneider \& Weiss, 1986;
Mao \& Paczynski, 1991; Udalski et al., 1993b). Indeed, the most intriguing of these
deviations is the one induced by a binary (or multiple) lens.

The study of light amplification curves produced by multiple lenses has not
yet been performed analytically because of the difficulties in the inversion
of the lens application. Anyway, these curves can be obtained numerically by
using some inversion algorithm (inverse ray shooting)(Wambsganns, 1997). From
the analytical point of view, only the caustics of a general binary lens
have been studied in some detail (Witt \& Petters, 1993). The lack of
analytical results utilizable in microlensing constitutes an irksome
obstacle in the complete interpretation of multiple microlensing events.

A particularly interesting case of multiple Schwarzschild lenses is formed
when one of the masses is much biggest than others (Mao \& Paczynski, 1991;
Gould \& Loeb, 1992). This is the situation of a typical planetary system
where a central star is surrounded by its planets bearing masses thousand or
million times smaller. The perturbations on Paczynski's curve induced by the
presence of a (even Earth - like) planet are in principle detectable by
collaboration teams exploiting world wide telescopic networks (Peale, 1997;
Sackett, 1997). Then microlensing could become a new efficient method for the
detection of small planets in extra - solar systems. This justifies the
major interest in this field that is growing in the last months. Preliminary
calculations on the probability of detection of planets have been made
(Gould \& Loeb, 1992; Bolatto \& Falco, 1994) and great efforts are lavished on the
problem of extraction of planetary parameters by approximate models (Gaudi \& Gould, 1996).
It is easy to imagine how
the availability of an analytical expression for light curves could help the
researches in this field.

The aim of this work is to describe planetary effects perturbatively,
exploiting the very little ratios of the masses of the planets with respect
to the star mass (Gould \& Loeb in 1992 first pioneered this kind of approach).
In section 2 the lens equation and other usual objects are
specified for the case of planetary systems. In section 3, by means of
perturbative theory, I derive the complete structure of critical curves and
caustics of a general (not only binary) planetary system; position of
planetary caustics and cusps are also found. In section 4 the problem of the
inversion of lens application is faced and resolved; consequently,
analytical microlensing light curves for planetary events thus obtained are
shown.

\begin{figure}
 \resizebox{\hsize}{!}{\includegraphics{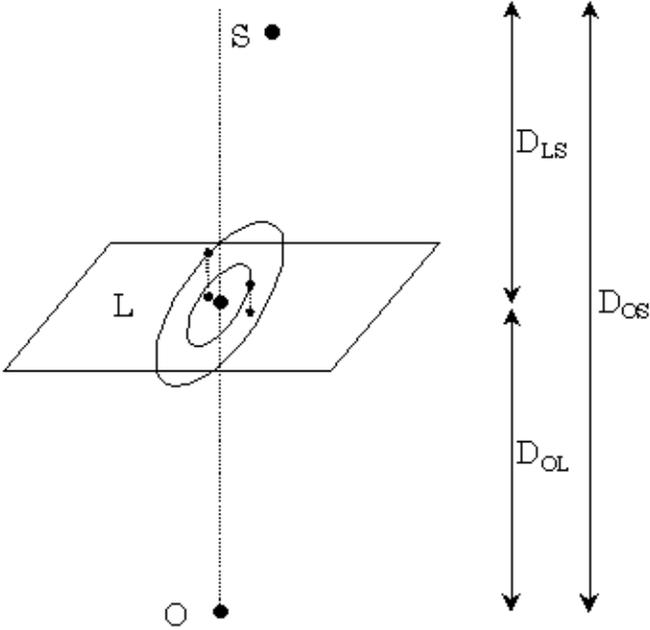}}
 \caption{Tipical scheme of planetary lensing.}
 \label{Fig general lensing}
\end{figure}

\section{Planetary lensing}

\noindent A planetary system is nothing but a discrete set of Schwarzschild
lenses in a small portion of the space. Fig. \ref{Fig general lensing} shows
the typical situation of planetary lensing. The lens plane is defined as a
plane orthogonal to the line of sight situated on the barycentre of the lens
mass distribution. According to the ordinary theory of lensing
(Schneider, Ehlers \& Falco, 1992), if the scale of this distribution is much smaller than
the distances separating the lens from the source and the observer, then one
can deal with the density projected on the lens plane instead of considering
the original volume density. This hypothesis is quite verified in real
observations where the typical distances are at least of the order of kpc.
This consideration can be important in planetary systems where very far
planets can eventually have projections enough near to the central star to
give perturbations comparable to those of planets placed in more favourable
positions. This means that multiple events could be less out of common than
one could think (Wambsganns, 1997; Gaudi, Naber \& Sackett, 1998). So, it is desirable
to preserve the whole planetary system as much as possible before abandoning
it for the simplest case of the single planet around the big star. We shall
see that the perturbative theory has the considerable advantage of being
rather insensitive to the number of planets as regards the difficulty of the
problem.

Let's define the length $R_{E}^{\odot }=\sqrt{\frac{4GM_{\odot }}{c^{2}}%
\frac{D_{LS}D_{OL}}{D_{OS}}}$. $\mathbf{x}=\left( x_{1};x_{2}\right) $
shall denote the coordinates in the lens plane normalized to $R_{E}^{\odot }$%
, while $\mathbf{y}=\left( y_{1};y_{2}\right) $ shall be the coordinates
in the source plane normalized to $R_{E}^{\odot }\frac{D_{OS}}{D_{OL}}$. $%
m_{1}$ will be the mass of the central star and $m_{2}$, ..., $m_{n}$ will
be the masses of the planets. All of these masses are meant to be measured
in solar masses. The star will always be placed at the origin, while the
projection on the lens plane of the i$_{th}$ planet will be denoted by $%
\mathbf{x}_{i}=\left( x_{i1};x_{i2}\right) $. With these notations, the
lens equation reads:
\begin{equation}
\mathbf{y}=\mathbf{x}-\frac{m_{1}\mathbf{x}}{\left| \mathbf{x}%
\right| ^{2}}-\sum\limits_{i=2}^{n}\frac{m_{i}\left( \mathbf{x}-%
\mathbf{x}_{i}\right) }{\left| \mathbf{x}-\mathbf{x}_{i}\right| ^{2}%
}  \label{General lens equation}
\end{equation}

In (\ref{General lens equation}), the deviation of light rays due to the
star has been explicitly separated by those of the planets. Given a source
position $\mathbf{y}$, the corresponding $\mathbf{x}$ solving the lens
equation are called images.

Many interesting properties of this vectorial application can be studied
through its Jacobian matrix. In particular, the determinant of this matrix
contains nearly all the information about the properties of the images.
Let's write the Jacobian determinant for the case of planetary lensing using
(\ref{General lens equation}):
\begin{multline}
\det J=1-\left[ \frac{m_{1}\left( x_{1}^{2}-x_{2}^{2}\right) }{\left(
x_{1}^{2}+x_{2}^{2}\right) ^{2}}+\sum\limits_{i=2}^{n}\frac{m_{i}\left(
\Delta _{i1}^{2}-\Delta _{i2}^{2}\right) }{\left( \Delta _{i1}^{2}+\Delta
_{i2}^{2}\right) ^{2}}\right] ^{2}+  \label{General Jacobian} \\
-4\left[ \frac{m_{1}x_{1}x_{2}}{\left( x_{1}^{2}+x_{2}^{2}\right) ^{2}}%
+\sum\limits_{i=2}^{n}\frac{m_{i}\Delta _{i1}\Delta _{i2}}{\left( \Delta
_{i1}^{2}+\Delta _{i2}^{2}\right) ^{2}}\right] ^{2}
\end{multline}

where $\mathbf{\Delta }_{i}=\left( \Delta _{i1};\Delta _{i2}\right) =%
\mathbf{x}-\mathbf{x}_{i}.$ Given an image I at position $\mathbf{x}%
_{I}$, the sign of $\det J\left( \mathbf{x}_{I}\right) $ is called the
parity of I. It can be proved that the amplification of the image I is given
by
\begin{equation}
\mu _{I}=\frac{1}{\left| \det J\left( \mathbf{x}_{I}\right) \right| }
\label{General amplification}
\end{equation}

We see from this equation that when $\det J$ is null, the amplification
diverges. This is rigorously true for point sources (in ray optics), while for finite (real)
sources the integration over the source's surface makes the amplification
finite (Witt \& Mao, 1994). The points where $\det J$ vanishes are called
critical points, the corresponding points in the source plane through (\ref
{General lens equation}) constitute the caustics. So, a point source
crossing a caustic will produce images with infinite amplification. Real
sources crossing caustics by some of their parts will be highly (but not
infinitely) amplified.

The structure of critical curves and caustics provides a substantial
description of the general behaviour of the lens (number and roughly
location of the images can be established). In microlensing, it is possible
to give a qualitative description of light curves just checking whether the
track of the source threads some caustic or not.

\begin{figure}
 \resizebox{\hsize}{!}{\includegraphics{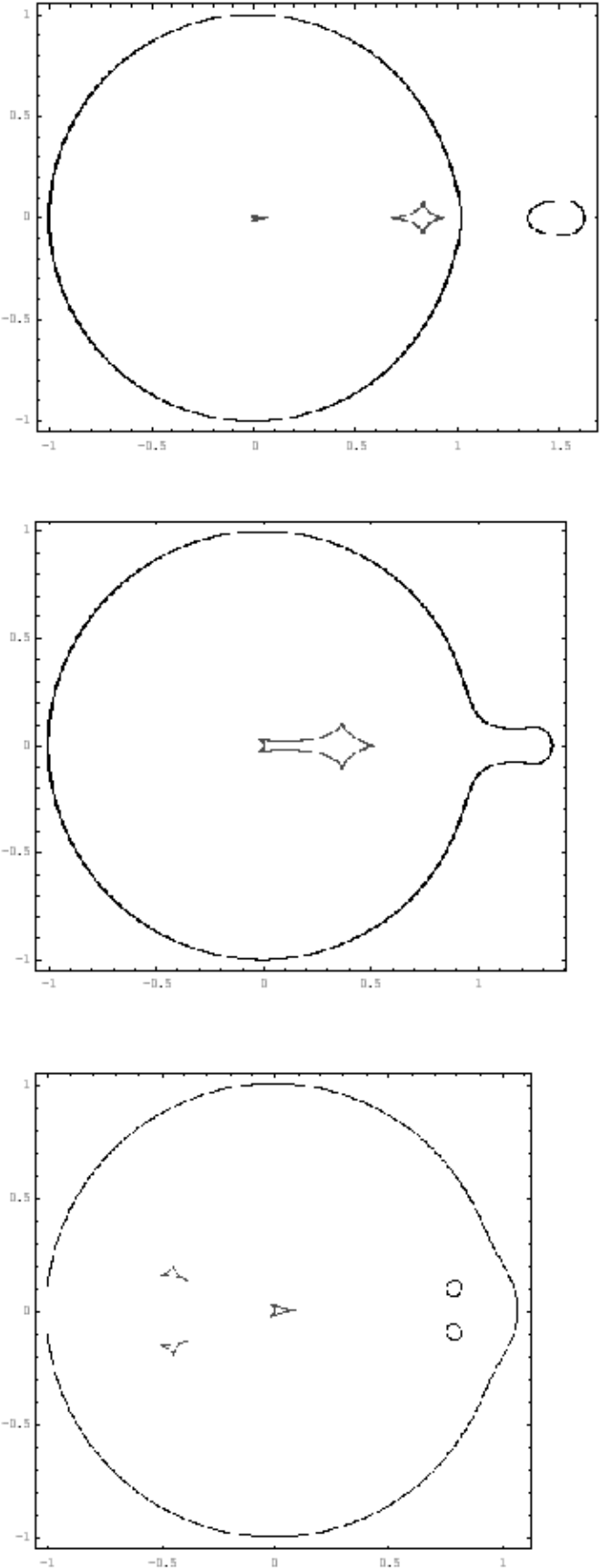}}
 \caption{There are three
possible situations in planetary lensing as regards critical
curves, shown in these plots. The smooth round curves are critical
curves, while the small cusped curves are the corresponding
caustics.}
 \label{Fig general critical curves}
\end{figure}

In fig. \ref{Fig general critical curves} the (numerically obtained)
critical curves and the caustics of a star with a single planet are shown.
There are three possible cases in such a situation. I recall that for a
single point source the critical curve is a ring with radius given by its
Einstein radius $\sqrt{m}$, while the caustic is a point in the origin of
the source plane. When the planet is far beyond the star's Einstein ring,
there is only a small perturbation of the two rings which lends finite
extension to the originally point - like caustics. This is much more evident
in the planetary caustic which is also displaced towards the star. When the
planet is in the proximity of the star's Einstein ring, the two critical
curves merge and so do the caustics. In the last situation where the planet
is internal to the star's Einstein ring, the star's critical curve returns
to be very near to a ring while the planet's critical curve turns into two
ovals to which a couple of triangular caustics correspond behind the star.

\section{Caustics and perturbative analysis}

\noindent In the solar system, the mass ratios between planets and the sun
are always less than one thousandth. Jupiter is $9.4\times 10^{-4}M_{\odot }$%
. Other planets are even less: Earth is $3\times 10^{-6}M_{\odot }$. With
these numbers, it is natural to expect that the presence of planets should
cause only little perturbations to the single lens case. Upon this
consideration the perturbative hypothesis is based. In this and the
following section the ratios between planetary and stellar masses will play
the role of perturbative expansions parameters. We shall see that in most
cases a first order expansion is sufficient to get very reliable results.

Let's turn to the caustics of planetary systems. First I shall examine the
modifications induced in the star's Einstein ring and consequently the
central caustic. Then I shall deal with planetary caustics.

\subsection{Central caustic}

\begin{figure*}
 \resizebox{12cm}{!}{\includegraphics{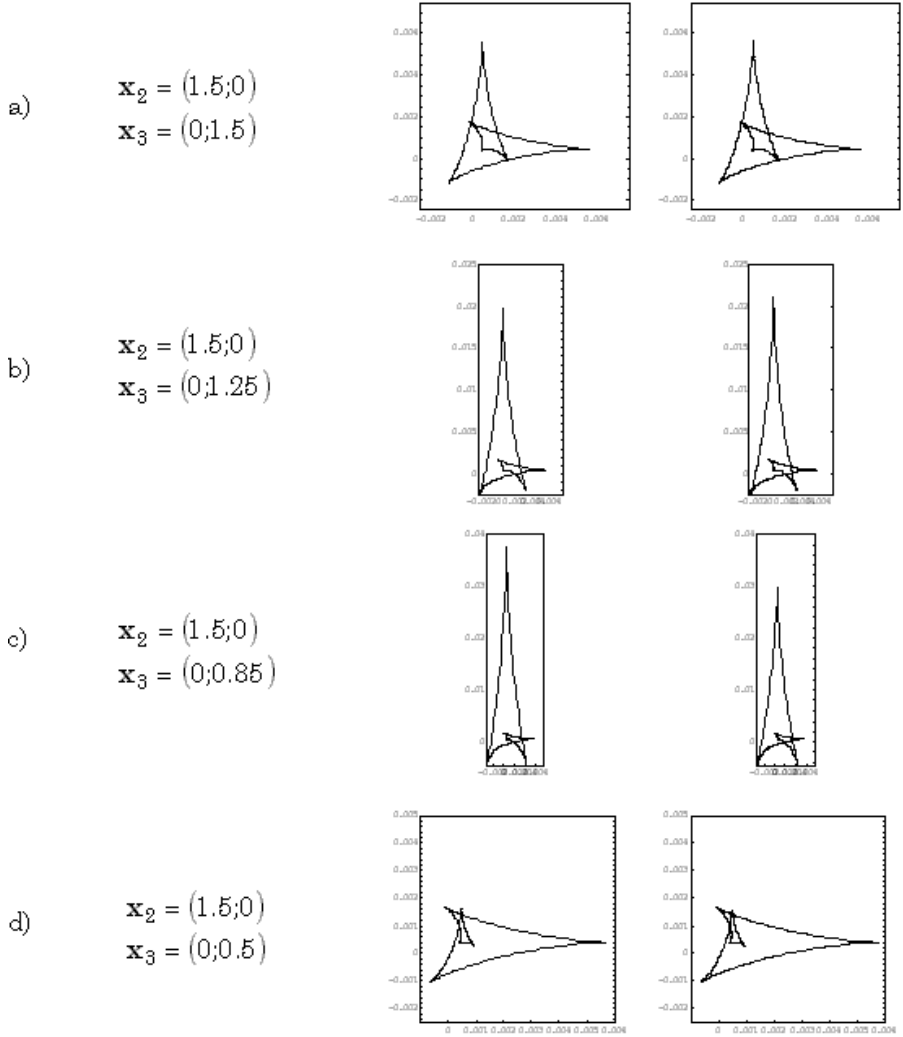}}
 \hfill
 \parbox[b]{55mm}{
 \caption{Central caustic for
Jovian planets. The caustics on the left are those found
perturbatively, while those on the right are obtained numerically.
Little differences occur when one of the planets is close to the
star's Einstein ring.}
 \label{Fig central caustic (Jupiter)}}
\end{figure*}

Of course, the starting point for the study of critical curves is the
equation $\det J=0$, which can easily be written in polar coordinates:
\begin{multline}
\left[ \frac{m_{1}\left( \cos ^{2}\vartheta -\sin ^{2}\vartheta \right) }{%
r^{2}}+\sum\limits_{i=2}^{n}\frac{m_{i}\left( \Delta _{i1}^{2}-\Delta
_{i2}^{2}\right) }{\left( \Delta _{i1}^{2}+\Delta _{i2}^{2}\right) ^{2}}%
\right] ^{2}+ \\
+4\left[ \frac{m_{1}\sin \vartheta \cos \vartheta }{r^{2}}%
+\sum\limits_{i=2}^{n}\frac{m_{i}\Delta _{i1}\Delta _{i2}}{\left( \Delta
_{i1}^{2}+\Delta _{i2}^{2}\right) ^{2}}\right] ^{2}=1
\end{multline}

Here $\mathbf{\Delta }_{i}=\left( r\cos \vartheta -x_{i1};r\sin \vartheta
-x_{i2}\right) $. Expanding this equation to the first order in $m_{i}$, we
get:
\begin{multline}
\frac{m_{1}^{2}}{r^{4}}+\frac{2m_{1}\left( \cos ^{2}\vartheta -\sin
^{2}\vartheta \right) }{r^{2}}\sum\limits_{i=2}^{n}\frac{m_{i}\left( \Delta
_{i1}^{2}-\Delta _{i2}^{2}\right) }{\left( \Delta _{i1}^{2}+\Delta
_{i2}^{2}\right) ^{2}}+  \label{Central detJ=0} \\
+\frac{8m_{1}\sin \vartheta \cos \vartheta }{r^{2}}\sum\limits_{i=2}^{n}%
\frac{m_{i}\Delta _{i1}\Delta _{i2}}{\left( \Delta _{i1}^{2}+\Delta
_{i2}^{2}\right) ^{2}}=1
\end{multline}

The zeroth order solution is simply $r=\sqrt{m_{1}}$, i.e. the Einstein
ring. Let's write the first order solution as:
\begin{equation}
r=\sqrt{m_{1}}\left( 1+\varepsilon \left( \vartheta \right) \right)
\end{equation}
with $\varepsilon \ll 1$. Substituting in (\ref{Central detJ=0}) and
expanding to the first order in $\varepsilon $. The zeroth order solution
cancels and $\varepsilon $ is found solving the remaining first degree
equation:
\begin{multline}
\varepsilon \left( \vartheta \right) =\frac{1}{2}\left( \cos ^{2}\vartheta
-\sin ^{2}\vartheta \right) \sum\limits_{i=2}^{n}\frac{m_{i}\left[ \left(
\Delta _{i1}^{0}\right) ^{2}-\left( \Delta _{i2}^{0}\right) ^{2}\right] }{%
\left[ \left( \Delta _{i1}^{0}\right) ^{2}+\left( \Delta _{i2}^{0}\right)
^{2}\right] ^{2}}+  \label{Central critic curve} \\
+2\sin \vartheta \cos \vartheta \sum\limits_{i=2}^{n}\frac{m_{i}\Delta
_{i1}^{0}\Delta _{i2}^{0}}{\left[ \left( \Delta _{i1}^{0}\right) ^{2}+\left(
\Delta _{i2}^{0}\right) ^{2}\right] ^{2}}
\end{multline}
where $\mathbf{\Delta }_{i}^{0}=\left( \sqrt{m_{1}}\cos \vartheta -x_{i1};%
\sqrt{m_{1}}\sin \vartheta -x_{i2}\right) $ is $\mathbf{\Delta }_{i}$ to
the zeroth order.

By very few steps we have found the perturbation of the Einstein ring in a
very simple way. The parametric equation of the central caustic is soon
found by applying the lens equation (\ref{General lens equation}) and
expanding again to the first order in $m_{i}$:
\begin{mathletters}
\label{Central caustic}
\begin{align}
y_{1}\left( \vartheta \right) & =2\sqrt{m_{1}}\varepsilon \left( \vartheta
\right) \cos \vartheta -\sum\limits_{i=2}^{n}\frac{m_{i}\Delta _{i1}^{0}}{%
\left[ \left( \Delta _{i1}^{0}\right) ^{2}+\left( \Delta _{i2}^{0}\right)
^{2}\right] } \\
y_{2}\left( \vartheta \right) & =2\sqrt{m_{1}}\varepsilon \left( \vartheta
\right) \sin \vartheta -\sum\limits_{i=2}^{n}\frac{m_{i}\Delta _{i2}^{0}}{%
\left[ \left( \Delta _{i1}^{0}\right) ^{2}+\left( \Delta _{i2}^{0}\right)
^{2}\right] }
\end{align}
\end{mathletters}

Of course, perturbative results are characterized by precise
limits of validity. In our case we see that when the denominators
in (\ref{Central critic curve}) vanish the perturbation diverges.
This is not allowed by our assumption that $\varepsilon $ must be
very small with respect to the unity. Those denominators represent
the distance between the planet and the general point of the
unperturbed star's Einstein ring. So we expect the perturbative
theory to fail in those portions of the critical curve which are
very near to one of the planets at least. We can understand this
``failure'' if we look back at fig. \ref{Fig general critical
curves}b: when the planet is close to the star's Einstein ring,
there is only one critical curve which is somewhat different from
the ring in the proximity of the planet. For some values of
$\vartheta $, the radial coordinate describing the critical curve
assumes also more than one value; this situation cannot be
described by a first order approximation, where, as we saw, the
perturbation solves a first degree equation.

The most interesting aspect of eqs. (\ref{Central critic curve}) and (\ref
{Central caustic}) is that they are comprehensive of the action of the whole
planetary system: they are valid for an arbitrary number of planets, not only
the classically investigated case of the single planet. So these formulas
enjoy a very high generality and can be used in more realistic contexts. We
also note that the contributions coming from different planets superpose
without interfering. This is an obvious consequence of the first order
approximation; if I had included second order terms, I would have found
``interaction'' between planets. These ``interaction'' terms are thus not
relevant in a first approximation.

Now let's compare the perturbative caustics with those found by
classical numerical algorithms to test the validity of the
perturbative approach. In fig. \ref{Fig central caustic (Jupiter)}
I show the results for the case of two Jovian planets placed in
several positions. When the planets are far enough from the
Einstein ring (fig. \ref{Fig central caustic (Jupiter)}a, \ref{Fig
central caustic (Jupiter)}d), the caustic found according to (\ref
{Central caustic}) is entirely identical to the numerical one.
Letting one of the planets approach the Einstein ring, a small
deviation begins appearing in the region coming from the portion
of the Einstein ring that is closest to the planet. This deviation
manifests itself in the size of the largest cusp. For Jovian
planets these discrepancies unveil at distances from the star's
Einstein ring of the order of a tenth of the Einstein radius.
These first encouraging results become much better in the case of
Earth - like planets. We expect the range of validity of
perturbative results to be increased for this kind of planets,
because of their smaller mass. Fig. \ref{Fig central caustic
(Earth)} shows that for these little planets things go very well
down to a hundredth of the Einstein radius.

\begin{figure}
 \resizebox{\hsize}{!}{\includegraphics{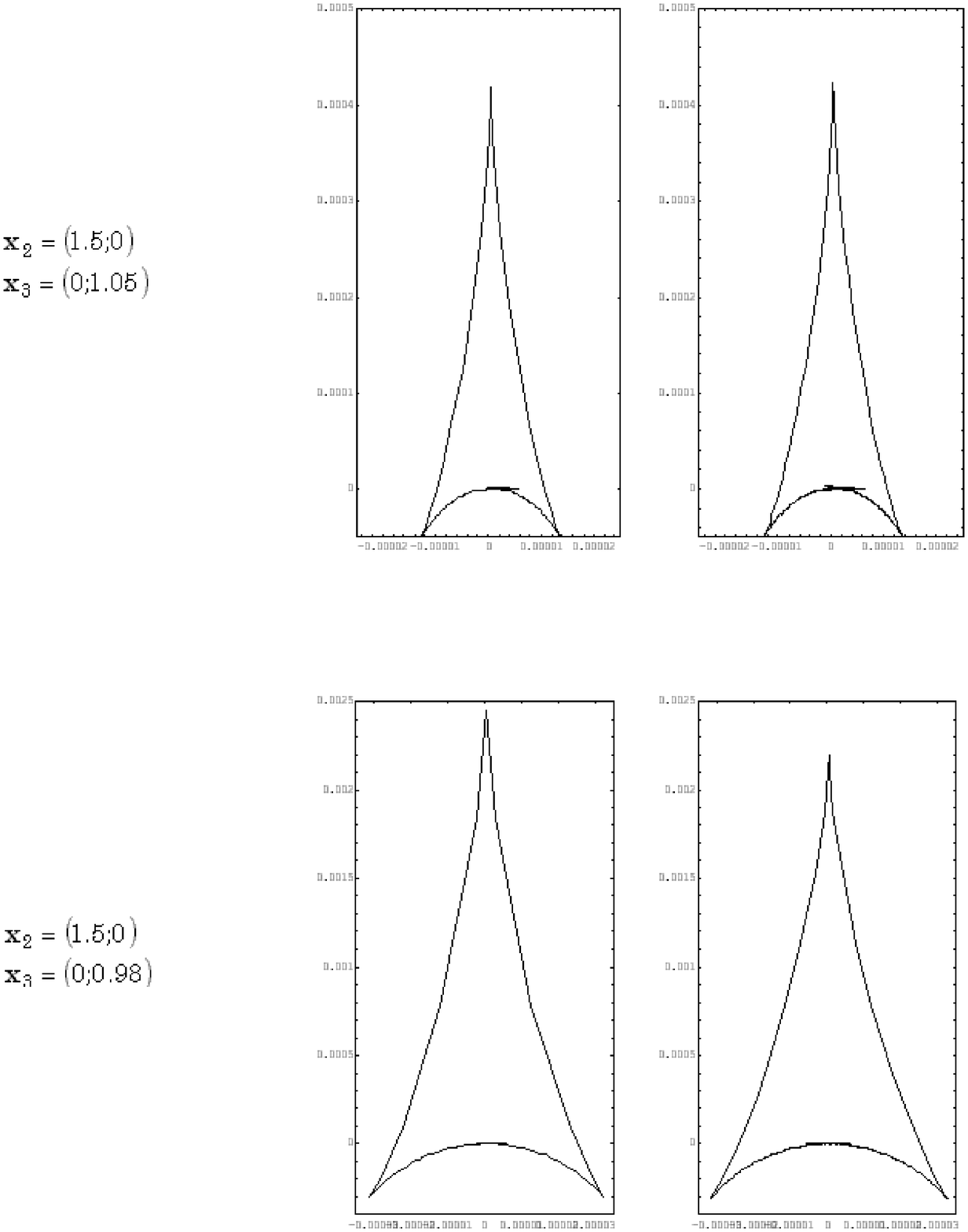}}
 \caption{Central caustics for Earth
- like planets. The perturbative ones are on the left and the
numerical on the right.}
 \label{Fig central caustic (Earth)}
\end{figure}

So the perturbative method is likely to provide reliable results at the
first order already. Moreover, it is not to be forgotten that, in principle,
the approximations can be improved pushing farther the perturbative
expansion.

\subsection{Planetary caustics}

Planetary caustics are usually studied considering the planet as a point-lens with
an external shear due to the star's gravitational field (Schneider, Ehlers \& Falco, 1992).
This kind of lens was introduced by Chang \& Refsdal (1979; 1984) in a cosmological context.
However, this lens is valid in planetary systems only to the lowest non-trivial
order in $m_{i}$. Therefore a correct study should only retain the lowest order terms
in the critical curves equation, so that Chang \& Refsdal's caustics are a suitable approximation
at the lowest order only and not beyond.

In order to complete the discussion of caustics in planetary systems and study their
features properly, in this subsection I derive planetary caustics from perturbative hypothesis
paying full attention to the order of each term.
The situation for planetary caustics is rather different from that of the
central caustic. There is no zeroth order solution to start from, since
their very presence is perturbative. Nevertheless this is not a great
problem: in fact we shall just search for the lowest order solution of the
critical curves equation.

To achieve this, I now rewrite $\det J=0$ in polar coordinates choosing the
planet $m_{2}$ situated in $\mathbf{x}_{2}$ as the origin:

\begin{multline}
\left[ \frac{m_{1}\left( \Delta _{11}^{2}-\Delta _{12}^{2}\right) }{\left(
\Delta _{11}^{2}+\Delta _{12}^{2}\right) ^{2}}+\frac{m_{2}\left( \cos
^{2}\vartheta -\sin ^{2}\vartheta \right) }{r^{2}}+ \right.
\label{Planetary detJ=0} \\
\left. + \sum\limits_{i=3}^{n}%
\frac{m_{i}\left( \Delta _{i1}^{2}-\Delta _{i2}^{2}\right) }{\left( \Delta
_{i1}^{2}+\Delta _{i2}^{2}\right) ^{2}}\right] ^{2}+
4\left[ \frac{m_{1}\Delta _{11}\Delta _{12}}{\left( \Delta _{11}^{2}+\Delta
_{12}^{2}\right) ^{2}}+ \right. \\
\left. + \frac{m_{2}\sin \vartheta \cos \vartheta }{r^{2}}%
+\sum\limits_{i=3}^{n}\frac{m_{i}\Delta _{i1}\Delta _{i2}}{\left( \Delta
_{i1}^{2}+\Delta _{i2}^{2}\right) ^{2}}\right] ^{2}=1
\end{multline}

When the planet is very far from the star, we know that its critical curve
tends to an Einstein ring with radius $r=\sqrt{m_{2}}$. So we search for
critical curves solving (\ref{Planetary detJ=0}) with $r\sim \sqrt{m_{2}}$.
Let's save the lowest order terms only. In this operation, the contributions
coming from the other planets are ejected out from the equation. It is
convenient to place the star in the usual position $(-x_{p};0)$. What
remains is:
\begin{equation}
\frac{m_{1}^{2}}{x_{p}^{4}}+\frac{m_{2}^{2}}{r^{4}}+\frac{2m_{1}m_{2}}{%
x_{p}^{2}r^{2}}\left( \cos ^{2}\vartheta -\sin ^{2}\vartheta \right) =1
\end{equation}
which is biquadratic in $r$. The solution is:
\begin{equation}
r=\sqrt{m_{2}\frac{\frac{m_{1}}{x_{p}^{2}}\cos 2\vartheta \pm \sqrt{1-\frac{%
m_{1}^{2}}{x_{p}^{4}}\sin ^{2}2\vartheta }}{\left( 1-\frac{m_{1}^{2}}{%
x_{p}^{4}}\right) }}  \label{Planetary critical curve}
\end{equation}
which verifies our assumption $r\sim \sqrt{m_{2}}$. The parametric equations
of caustics can be found in the usual way substituting (\ref{Planetary
critical curve}) in the lens application and expanding to the first non
trivial order ($\sqrt{m_{2}}$):
\begin{mathletters}
\label{Planetary caustics}
\begin{eqnarray}
y_{1} &=&x_{p}-\frac{m_{1}}{x_{p}}\left( 1-\frac{r}{x_{p}}\cos \vartheta
\right) +\left( r-\frac{m_{2}}{r}\right) \cos \vartheta \\
y_{2} &=&-\frac{m_{1}}{x_{p}^{2}}r\sin \vartheta +\left( r-\frac{m_{2}}{r}%
\right) \sin \vartheta
\end{eqnarray}
\end{mathletters}

The contributions from the other planets are again of higher order. So the
structure of planetary caustics is not affected by the presence of other
planets at the lowest order in a perturbative expansion. These formulas can
thus be used in a single planet situation as well as in a rich planetary
system.

Observe that $r$ goes to infinity as $x_{p}$ tends to $\sqrt{m_{1}}$, i.e.
when the planet is next to the star's Einstein radius. So the perturbative
results will not be valid in this situation. The reason is the same
discussed for the central caustic. The merging of the two caustic is not
describable in the lowest order perturbative expansion. Moreover, there's
another limit to be taken in account. I have eliminated all the terms coming
from the other planets because of their higher order. But these terms can
become dominant when their denominators are small. This happens when one of
these planets is close to the planet we are examining. This is not an exotic
situation since we must always remember that what counts is actually the
projection of the positions on the lens plane. So planets could be very far
apart but have near projections yielding exotic critical curves.

We see that the critical curves traced by (\ref{Planetary critical curve})
have two branches according to the double sign in their expression. For
planets external to the star's Einstein ring ( $x_{p}>\sqrt{m_{1}}$), the
branch coming from the positive sign is real while the other coming from the
negative sign is imaginary for all values of $\vartheta $, being $\left|
\frac{m_{1}}{x_{p}^{2}}\cos 2\vartheta \right| <\sqrt{1-\frac{m_{1}^{2}}{%
x_{p}^{4}}\sin ^{2}2\vartheta }$. For internal planets ( $x_{p}<\sqrt{m_{1}}$%
), the denominator is negative and $\left| \frac{m_{1}}{x_{p}^{2}}\cos
2\vartheta \right| >\sqrt{1-\frac{m_{1}^{2}}{x_{p}^{4}}\sin ^{2}2\vartheta }$%
. So the two branches are both real for:
\begin{equation}
\sin ^{2}2\vartheta <\frac{x_{p}^{4}}{m_{1}^{2}};\text{ \ }\cos 2\vartheta <0
\end{equation}
that is in two small regions around $\vartheta =\frac{\pi }{2}$ and $%
\vartheta =\frac{3\pi }{2}$. They are both imaginary elsewhere. All these
results are coherent with the behaviour exposed in fig. \ref{Fig general
critical curves}. We have one planetary caustic for external planets and two
disconnected caustics for internal planets.

\begin{figure*}
 \resizebox{12cm}{!}{\includegraphics{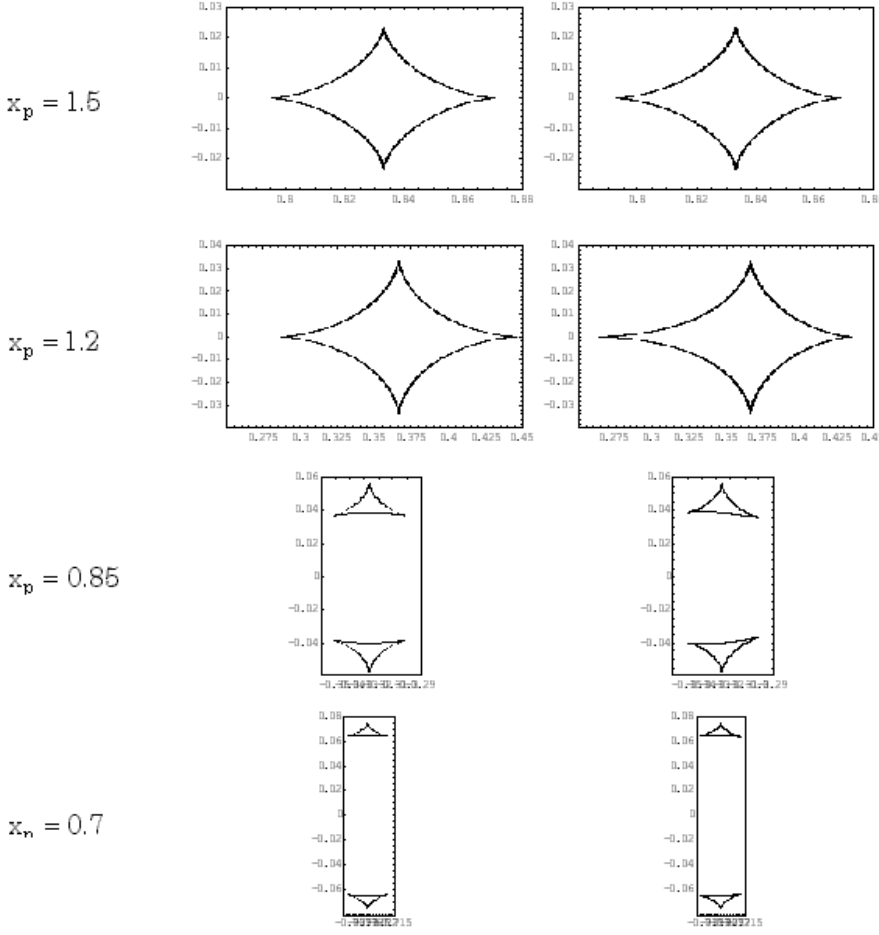}}
 \hfill
 \parbox[b]{55mm}{
 \caption{Planetary caustics for a
Jovian planet. The caustics on the left column are perturbative
while the ones on the right are numerical.}
 \label{Fig planetary caustics}}
\end{figure*}

Fig. \ref{Fig planetary caustics} shows the comparison with the numerical caustics.
In fig. \ref{Fig planetary caustics}b the discrepancy with
the numerical results appears as a loss of symmetry of the numerical caustic
which is elongated towards the central star. This effect is not present in
the perturbative Chang \& Refsdal's one. For internal planets near the star's Einstein ring,
the basis of the triangular perturbative caustic is parallel to the
star - planet axis(fig. \ref{Fig planetary
caustics}c). So Chang \& Refsdal's lens works good until the field can be taken as
uniform. When the spherical symmetry becomes important, the caustics begin to differ
from perturbative ones. These effects can be taken into account by considering higher
order terms in the expansion. These terms would provide the right corrections to the
Chang \& Refsdal's approximation.

Eqs. (\ref{Planetary critical curve}) and (\ref{Planetary caustics}) can be
employed to find interesting characteristics of planetary caustics. For
example, let's find the position of the couple of caustics for internal
planets. We saw that the critical curves are centered upon $\vartheta =\frac{%
\pi }{2}$ and $\vartheta =\frac{3\pi }{2}$. Consider the first of these (the
other is similar at all). Inserting these values of $\vartheta $ in (\ref
{Planetary critical curve}), the possible values of $r$ are:
\begin{equation}
r=\sqrt{m_{2}\frac{-\frac{m_{1}}{x_{p}^{2}}\pm 1}{\left( 1-\frac{m_{1}^{2}}{%
x_{p}^{4}}\right) }}
\end{equation}

The point
\begin{equation}
r=x_{p}\sqrt{\frac{m_{1}m_{2}}{m_{1}^{2}-x_{p}^{4}}}
\end{equation}
obtainable by a quadratic mean from the two values, is internal to the
critical curve and gives an approximation for its position. Immediately,
using the lens equation and expanding to the lowest order terms, we find the
position of the caustics:
\begin{mathletters}
\label{Planetary caustics position}
\begin{eqnarray}
y_{1} &=&x_{p}-\frac{m_{1}}{x_{p}} \\
y_{2} &=&x_{p}\sqrt{\frac{m_{1}m_{2}}{m_{1}^{2}-x_{p}^{4}}}\left[ 1-\frac{%
m_{1}}{x_{p}^{2}}-\frac{m_{1}^{2}-x_{p}^{4}}{m_{1}x_{p}^{2}}\right]
\end{eqnarray}
\end{mathletters}

\begin{figure}
 \resizebox{\hsize}{!}{\includegraphics{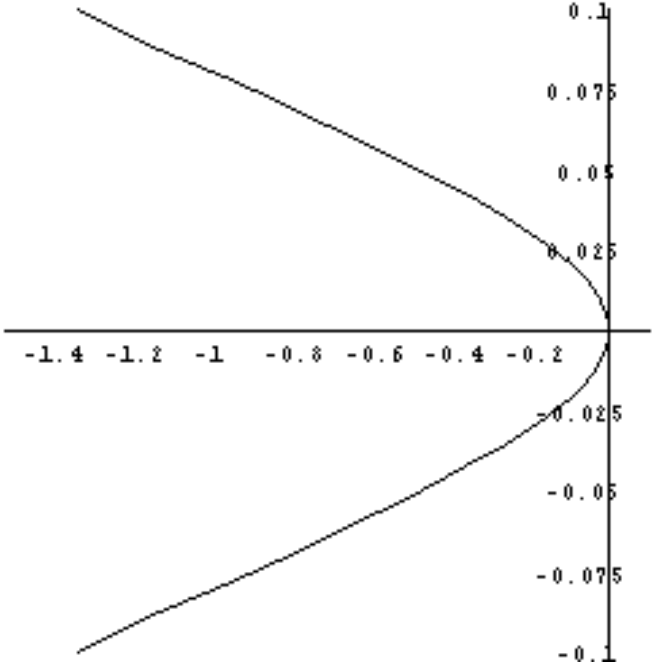}}
 \caption{For planets internal to the
star's Einstein ring, the planetary caustics move on this curve.}
 \label{Fig caustics position}
\end{figure}

The first of these is a well known formula (Griest \& Safizadeh, 1997). The
second completes the information given from the first and allows a complete
individuation of the two caustics. Fig. \ref{Fig caustics position} is a
plot of the position of the two caustics as a function of the distance of
the planet from the star. When $x_{p}\rightarrow 0$ the caustics
approximately move on the lines:
\begin{equation}
y_{2}=\pm 2\sqrt{\frac{m_{2}}{m_{1}}}y_{1}
\end{equation}

The two planetary caustics delimitate a region of high de - amplification
which can appear in microlensing light curves as negative peaks. The
positions of the caustics can give a measure of the size of this region and
consequently the size of these negative peaks.

\subsection{Cusps}

This subsection concludes the study of the perturbative caustics with the
analysis of the position of cusps in these caustics. The position of cusps
can be important in several studies such as microlensing itself. In fact
cusps are surrounded by a region with an amplification even higher than that
of fold singularities. They also define the extension and the shape of the
caustic.

Cusps are defined as the points where the tangent vector of the caustic
vanishes. In order to find them we must set
\begin{equation}
\left\{
\begin{array}{c}
y_{1}^{\prime }\left( \vartheta \right) =0 \\
y_{2}^{\prime }\left( \vartheta \right) =0
\end{array}
\right.  \label{General cusp equation}
\end{equation}
and resolve this system of equations for $\vartheta $.

Let's start with the central caustic. Eqs. (\ref{General cusp equation})
after several steps become:
\begin{equation}
\left\{
\begin{array}{c}
\cos \vartheta \left(\frac{\partial \varepsilon }{\partial \vartheta }-
\sum\limits_{i=2}^{n}m_{i}\frac{\left( \Delta _{i2}^{0}\cos
\vartheta -\Delta _{i1}^{0}\sin \vartheta \right) \left( \Delta
_{i1}^{0}\cos \vartheta +\Delta _{i2}^{0}\sin \vartheta \right) }{\left[
\left( \Delta _{i1}^{0}\right) ^{2}+\left( \Delta _{i2}^{0}\right) ^{2}%
\right] ^{2}}\right)=0 \\
\sin \vartheta \left(\frac{\partial \varepsilon }{\partial \vartheta }-
\sum\limits_{i=2}^{n}m_{i}\frac{\left( \Delta _{i2}^{0}\cos
\vartheta -\Delta _{i1}^{0}\sin \vartheta \right) \left( \Delta
_{i1}^{0}\cos \vartheta +\Delta _{i2}^{0}\sin \vartheta \right) }{\left[
\left( \Delta _{i1}^{0}\right) ^{2}+\left( \Delta _{i2}^{0}\right) ^{2}%
\right] ^{2}}\right)=0
\end{array}
\right.
\end{equation}

These can simultaneously vanish only if
\begin{equation}
\frac{\partial \varepsilon }{\partial \vartheta }-\sum\limits_{i=2}^{n}m_{i}%
\frac{\left( \Delta _{i2}^{0}\cos \vartheta -\Delta _{i1}^{0}\sin \vartheta
\right) \left( \Delta _{i1}^{0}\cos \vartheta +\Delta _{i2}^{0}\sin
\vartheta \right) }{\left[ \left( \Delta _{i1}^{0}\right) ^{2}+\left( \Delta
_{i2}^{0}\right) ^{2}\right] ^{2}}=0
\end{equation}

Explicitly, this equation is:
\begin{multline}
\sum\limits_{i=2}^{n}\frac{m_{i}}{\left[ \left( \Delta _{i1}^{0}\right)
^{2}+\left( \Delta _{i2}^{0}\right) ^{2}\right] ^{3}}\left\{ 3\left[ \left(
\Delta _{i2}^{0}\right) ^{4}-\left( \Delta _{i1}^{0}\right) ^{4}\right] \sin
\vartheta \cos \vartheta +\right. \\
+3\Delta _{i1}^{0}\Delta _{i2}^{0}\left[ \left( \Delta _{i1}^{0}\right)
^{2}+\left( \Delta _{i2}^{0}\right) ^{2}\right] \left( \cos ^{2}\vartheta
-\sin ^{2}\vartheta \right) + \\
-\Delta _{i1}^{0}\sqrt{m_{1}}\left[ \left( \Delta _{i1}^{0}\right)
^{2}-3\left( \Delta _{i2}^{0}\right) ^{2}\right] \left( 4\sin ^{3}\vartheta
-3\sin \vartheta \right) + \\
\left. +\Delta _{i2}^{0}\sqrt{m_{1}}\left[ \left( \Delta _{i2}^{0}\right)
^{2}-3\left( \Delta _{i1}^{0}\right) ^{2}\right] \left( 4\cos ^{3}\vartheta
-3\cos \vartheta \right) \right\}
\end{multline}
which, in despite of its cumbersome aspect, can be exactly solved in the
case of the single planet where it yields the four solutions:
\begin{multline}
\vartheta =0\text{; \ }\vartheta =\pi \text{; \ } \\
\vartheta =\pm \arccos %
\left[ \frac{3m_{1}+3x_{p}^{2}-\sqrt{9m_{1}^{2}-14m_{1}x_{p}^{2}+9x_{p}^{4}}%
}{4\sqrt{m_{1}}x_{p}}\right]
\end{multline}

For planetary caustics, we can proceed in a similar way. Multiplying (\ref
{General cusp equation}a) by $\sin \vartheta $, (\ref{General cusp equation}%
b) by $\cos \vartheta $ and subtracting, we have:
\begin{equation}
\frac{m_{1}}{x_{p}^{2}}r\left( \cos ^{2}\vartheta -\sin ^{2}\vartheta
\right) +2\frac{m_{1}}{x_{p}^{2}}\frac{\partial r}{\partial \vartheta }\sin
\vartheta \cos \vartheta -r+\frac{m_{2}}{r}=0
\end{equation}

Multiplying by $r$, we get an equation in $r^{2}$ which is easier to handle:
\begin{equation}
\frac{m_{1}}{x_{p}^{2}}r^{2}\left( \cos ^{2}\vartheta -\sin ^{2}\vartheta
\right) +\frac{m_{1}}{x_{p}^{2}}\frac{\partial r^{2}}{\partial \vartheta }%
\sin \vartheta \cos \vartheta -r^{2}+m_{2}=0
\end{equation}

Inserting (\ref{Planetary critical curve}) and solving, we have:
\begin{equation}
\vartheta =0\text{; \ }\vartheta =\frac{\pi }{2}\text{; \ }\vartheta =\pi
\text{; \ }\vartheta =\frac{3\pi }{2}
\end{equation}
on the higher branch, and:
\begin{mathletters}
\begin{eqnarray}
\vartheta &=&\pm \frac{1}{2}\arcsin \left[ \frac{\sqrt{3}x_{p}^{2}}{2m_{1}}%
\right] \text{;} \\
\vartheta &=&\pi \pm \frac{1}{2}\arcsin \left[ \frac{\sqrt{3%
}x_{p}^{2}}{2m_{1}}\right] \text{;} \\
\vartheta &=&\frac{\pi }{2}\pm \frac{1}{2}\arcsin \left[ \frac{\sqrt{3}%
x_{p}^{2}}{2m_{1}}\right] \text{;} \label{Lower planetary cusps 1} \\
\vartheta &=&\frac{3\pi }{2}\pm \frac{1}{2%
}\arcsin \left[ \frac{\sqrt{3}x_{p}^{2}}{2m_{1}}\right]
\label{Lower planetary cusps 2}
\end{eqnarray}
\end{mathletters}
on the lower.

For external planets, the higher branch is complete while the lower is
absent, so only the four cusps on the higher branch are actually present.
For internal planets, the two branches are real only near $\vartheta =\frac{%
\pi }{2}$ and $\vartheta =\frac{3\pi }{2}$. So the higher branch has the two
cusps at $\vartheta =\frac{\pi }{2}$ and $\vartheta =\frac{3\pi }{2}$, while
in the lower one the four cusps (\ref{Lower planetary cusps 1}) and
(\ref{Lower planetary cusps 2}) are real and
the others are imaginary. Summing up we have six cusps distributed in such a
way as to form two triangular caustics.

\section{Microlensing}

\noindent Among the numerous forms of gravitational lensing, microlensing is
surely one of the most relevant since it opens the possibility of probing
the galactic structure through a directly gravitational investigation.

Microlensing occurs when the images of a given source, produced by a small
lens, are too close (typically less than $10^{-3}$ arcsecs) to be separated
by our telescopes. As we cannot see but a point image of the source, the
only way to notice a lensing effect is through a variation of the total
light flux coming from the observed source. For a point lens mass, this
variation was found by Paczynski (1986) who first thought of galactic
microlensing as a new observable astronomical phenomenon. For a planetary
system the anomalies in amplification patterns do not enjoy a full analytical
description. Our aim
is to use a perturbative approach to solve this problem and find analytical
light curves for stars accompanied by their planets.

\subsection{Paczynski's curve}

\begin{figure}
 \resizebox{\hsize}{!}{\includegraphics{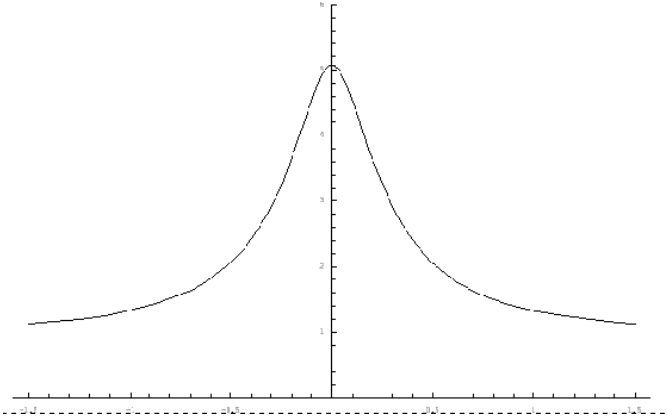}}
 \caption{Typical microlensing curve
for a point lens mass.}
 \label{Fig Paczynski's curve}
\end{figure}

Before considering the problem of planetary microlensing, it is
useful to review the steps to be followed in order to get
amplification light curves in the event of a single mass
(Schneider, Ehlers \& Falco, 1992). This will help us in fixing
the problems to be faced. In this case, the lens equation takes a
very simple aspect:
\begin{equation}
\mathbf{y}=\mathbf{x}-\frac{m_{1}}{\left| \mathbf{x}\right| ^{2}}%
\mathbf{x}  \label{Single lens equation}
\end{equation}

The total amplification is found by summing the amplification of all images.
So the first step is to find these images, i.e. the lens equation is to be
inverted. Here the task is quite easy, because (\ref{Single lens equation})
reduces to a second degree equation whose solutions are:
\begin{mathletters}
\label{Paczynski's images}
\begin{eqnarray}
\mathbf{I}^{+} &=&\frac{\mathbf{y}}{2}\left( 1+\frac{\sqrt{%
4m_{1}+\left| \mathbf{y}\right| ^{2}}}{\left| \mathbf{y}\right| }%
\right) \\
\mathbf{I}^{-} &=&\frac{\mathbf{y}}{2}\left( 1-\frac{\sqrt{%
4m_{1}+\left| \mathbf{y}\right| ^{2}}}{\left| \mathbf{y}\right| }%
\right)
\end{eqnarray}
\end{mathletters}

The next step is to compute the amplifications corresponding to each of
these images. According to (\ref{General amplification}), these are:
\begin{equation}
\mu _{I^{\pm }}=\frac{1}{\left| \det J\left( \mathbf{I}^{\pm }\right)
\right| }=\frac{1}{\left| 1-\frac{m_{1}^{2}}{\left| \mathbf{I}^{\pm }\right|
^{4}}\right| }
\end{equation}

It is interesting to study the properties of the two images to discover
their physical essence (Blandford \& Narayan, 1986). The image $\mathbf{I}%
^{+}$ has positive parity; in the limit of vanishing lensing effect $\left(
\left| \mathbf{y}\right| ^{2}\gg m_{1}\right) $ $\mathbf{I}^{+}$ tends
to $\mathbf{y}$ and its amplification becomes unitary. Thus $\mathbf{I}%
^{+}$ reduces to the usual image of the source in the absence of lensing. In
what follows I'll refer to it as the principal image. $\mathbf{I}^{-}$
has negative parity and in the limit of low lensing goes as $m_{1}/\left|
\mathbf{y}\right| $, while its amplification is always $\mu
_{I^{-}}=\mu _{I^{+}}-1$. I shall call it secondary image as it disappears
when the lensing effect is not present. Both images are aligned on the line
connecting the source and the lens: the principal image is always external
to the Einstein ring, while the secondary one is internal to it.

Now we have to sum up the two amplifications to obtain the so - called
amplification map:
\begin{equation}
\mu _{TOT}=\frac{2m_{1}+\left| \mathbf{y}\right| ^{2}}{\left| \mathbf{y%
}\right| \sqrt{4m_{1}+\left| \mathbf{y}\right| ^{2}}}
\label{Paczynski's amplification}
\end{equation}

This function tells us the amplification corresponding to any given position
of the source relatively to the lensing object. Of course it only depends on
the distance because of the symmetry of the lens.

The final step is to make the source move along a rectilinear trajectory to
obtain the complete light curves corresponding to the passage of a massive
lens near the line of sight of the source (obviously it makes no difference
who is moving, what counts is only the relative motion). The distance $%
\left| \mathbf{y}\right| $ is:
\begin{equation}
\left| \mathbf{y}\right| =\sqrt{b^{2}+v_{\bot }^{2}t^{2}}
\end{equation}
where $b$ is the impact parameter (the closest approach distance) and $%
v_{\bot }$ is the projection of the relative speed in a plane orthogonal to
the line of sight.

A typical light curve is shown in fig. \ref{Fig Paczynski's
curve}. The height of the maximum is found by substituting the
impact parameter in (\ref {Paczynski's amplification}). It becomes
infinite as $b$ tends to zero. Real sources have finite extensions
implying integration processes smoothing the peak of the curve
(Witt \& Mao, 1994). This cut - off becomes evident when $\left|
\mathbf{y}\right| $ is comparable to the source radius.

\subsection{The problem of planetary microlensing}

In principle, the procedure for attaining microlensing light curves for
multiple lenses is the same just expounded for a single point lens. First we
should invert the lens application, second we have to compute the
amplification of all the images, then sum up to have the amplification map
and finally introduce the motion of the source relatively to the lens
system. But if we write the lens equation for a star with just one planet
placed in $\mathbf{x}_{p}=\left( x_{p},0\right) $:
\begin{equation}
\mathbf{y}=\mathbf{x}-\frac{m_{1}}{\left| \mathbf{x}\right| ^{2}}%
\mathbf{x}-\frac{m_{2}}{\left| \mathbf{x}-\mathbf{x}_{p}\right| ^{2}%
}\left( \mathbf{x}-\mathbf{x}_{p}\right)
\label{One planet lens equation}
\end{equation}
we at once see that the inversion is not possible since one must surrender
at a fifth degree equation which doesn't allow to find the images produced
by such a lens.

\begin{figure}
 \resizebox{\hsize}{!}{\includegraphics{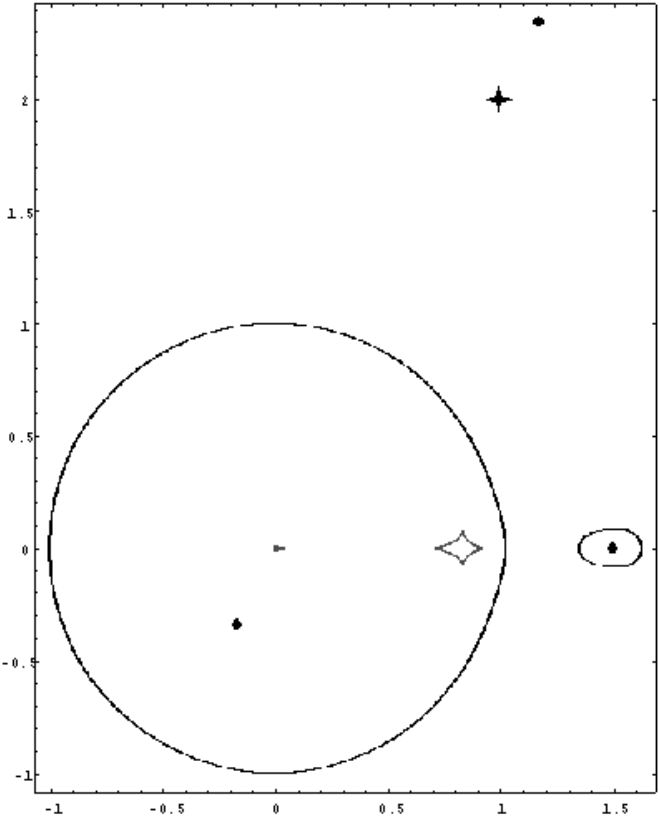}}
 \caption{In presence of a planet, a
source (here represented by a four-cornered star) placed outside
the caustics generates three images (the three little circles in
the figure).}
 \label{Fig planetary images}
\end{figure}

A glance at the numerical results can indicate us which way is to be taken
in the inversion of the lens application (\ref{One planet lens equation}).
When the source is outside the caustics, only three images are present (see
fig. \ref{Fig planetary images}). One of them is outside all critical curves
and approaches the source when the latter is far enough from the lens. This
is indeed the principal image. Another image is inside the star's critical
curve. It is easy to understand that when the mass of the planet vanishes
this image becomes the star's secondary image. The last image is near the
planet (inside the planetary critical curve when the planet is external to
the star's Einstein ring). I shall refer to this as the planetary
image. It is clear that the presence of the planet slightly perturbs the
principal and secondary image of the star, so that their position can be
found applying perturbation theory to Paczynski's images. The planetary
image is completely perturbative, since it is not present in the zeroth
order situation in which the planet is absent, and must be treated
separately. When the source threads a caustic, two new images are formed
with opposite parities whose effects are similar to those of the planetary
image.

So, the perturbative analysis is likely to be the key to solve the problem
of planetary microlensing. In the following two subsections I will use it to
discover the images and their amplification. Finally, I will build
amplification light curves and compare them with their numerical
counterparts.

\subsection{Principal and secondary image}

Paczynski's images (\ref{Paczynski's images}) are the starting point for our
expansion and will be generically indicated by the symbol $\mathbf{I}%
^{\left( 0\right) }$. Let's write the position of the images to the first
order in $m_{2}$ as the sum of Paczynski's image and a small perturbation
$\mathbf{\Delta I}$:
\begin{equation}
\mathbf{I}^{\left( 1\right) }=\mathbf{I}^{\left( 0\right) }+\mathbf{%
\Delta I}
\end{equation}
With this position, in the lens equation expanded to the first order in $%
m_{2}$%
\begin{mathletters}
\begin{eqnarray}
y_{1} =I_{1}^{\left( 1\right) }-\frac{m_{1}I_{1}^{\left( 1\right) }}{%
\left( I_{1}^{\left( 1\right) }\right) ^{2}+\left( I_{2}^{\left( 1\right)
}\right) ^{2}}+ \notag \\
-\frac{m_{2}\left( I_{1}^{\left( 0\right) }-x_{p}\right) }{%
\left( I_{1}^{\left( 0\right) }-x_{p}\right) ^{2}+\left( I_{2}^{\left(
0\right) }\right) ^{2}} \\
y_{2} =I_{2}^{\left( 1\right) }-\frac{m_{1}I_{2}^{\left( 1\right) }}{%
\left( I_{1}^{\left( 1\right) }\right) ^{2}+\left( I_{2}^{\left( 1\right)
}\right) ^{2}}+ \notag \\
-\frac{m_{2}I_{2}^{\left( 0\right) }}{\left( I_{1}^{\left(
0\right) }-x_{p}\right) ^{2}+\left( I_{2}^{\left( 0\right) }\right) ^{2}}
\end{eqnarray}
\end{mathletters}
the planetary term no longer contains the perturbation $\mathbf{\Delta I}$%
. Putting:
\begin{mathletters}
\begin{eqnarray}
\Delta y_{1}=\frac{m_{2}\left( I_{1}^{\left( 0\right) }-x_{p}\right) }{%
\left( I_{1}^{\left( 0\right) }-x_{p}\right) ^{2}+\left( I_{2}^{\left(
0\right) }\right) ^{2}}\text{;} \\
\Delta y_{2}=\frac{m_{2}I_{2}^{\left(
0\right) }}{\left( I_{1}^{\left( 0\right) }-x_{p}\right) ^{2}+\left(
I_{2}^{\left( 0\right) }\right) ^{2}}
\end{eqnarray}
\end{mathletters}
and bringing these terms to the left members, we re-gain the structure of
the Schwarzschild lens equation (\ref{Single lens equation}) in the variable
$\mathbf{I}^{\left( 1\right) }$ for the source position $\mathbf{y}+%
\mathbf{\Delta y}$. The planetary induced perturbation can be thus read
as a shift in the source position. $\mathbf{I}^{\left( 1\right) }$ has
the same expression as $\mathbf{I}^{\left( 0\right) }$ evaluated in $%
\mathbf{y}+\mathbf{\Delta y}$ instead of $\mathbf{y}$. The
perturbation $\mathbf{\Delta I}$ is found by expanding $\mathbf{I}%
^{\left( 1\right) }$ to the first order in $\mathbf{\Delta y}$:
\begin{mathletters}
\label{Images perturbation}
\begin{eqnarray}
\Delta I_{1}^{\pm } =\frac{\Delta y_{1}}{2}\left( 1\pm \frac{\sqrt{%
4m_{1}+y_{1}^{2}+y_{2}^{2}}}{\sqrt{y_{1}^{2}+y_{2}^{2}}}\right) \notag \\
\mp \frac{%
2m_{1}y_{1}\left( y_{1}\Delta y_{1}+y_{2}\Delta y_{2}\right) }{\sqrt{%
4m_{1}+y_{1}^{2}+y_{2}^{2}}\sqrt{\left( y_{1}^{2}+y_{2}^{2}\right) ^{3}}} \\
\Delta I_{2}^{\pm } =\frac{\Delta y_{2}}{2}\left( 1\pm \frac{\sqrt{%
4m_{1}+y_{1}^{2}+y_{2}^{2}}}{\sqrt{y_{1}^{2}+y_{2}^{2}}}\right) \notag \\
\mp \frac{%
2m_{1}y_{2}\left( y_{1}\Delta y_{1}+y_{2}\Delta y_{2}\right) }{\sqrt{%
4m_{1}+y_{1}^{2}+y_{2}^{2}}\sqrt{\left( y_{1}^{2}+y_{2}^{2}\right) ^{3}}}
\end{eqnarray}
\end{mathletters}

The upper signs stand for the principal image and the lower for the
secondary. The expansion parameter $m_{2}$ appears through $\mathbf{%
\Delta y}$.

Now the position of the principal and secondary image are known. The most
delicate operation is done and the door to the planetary microlensing is
open at last. What remains is only mechanic computation without any
conceptual difficulties.

The amplification of each image is found by the general formula (\ref
{General amplification}) expanded to the first order in $m_{2}$ (I drop the
zero from $\mathbf{I}^{\left( 0\right) }$ to simplify notation):
\begin{multline}
\mu _{I}=\frac{1}{\left| 1-\frac{m_{1}^{2}}{\left( I_{1}^{2}+I_{2}^{2}\right)
^{2}}\right| }\left\{ 1-\frac{4m_{1}^{2}}{\left( I_{1}^{2}+I_{2}^{2}\right)
^{3}}\left( I_{1}\Delta I_{1}+I_{2}\Delta I_{2}\right) +\right.
\label{Image amplification} \\
+\frac{2m_{1}m_{2}}{\left( 1-\frac{m_{1}^{2}}{\left(
I_{1}^{2}+I_{2}^{2}\right) ^{2}}\right) } \\
\left. \left[ \frac{\left(
I_{1}^{2}-I_{2}^{2}\right) \left( \left( I_{1}-x_{p}\right)
^{2}-I_{2}^{2}\right) +4I_{1}\left( I_{1}-x_{p}\right) I_{2}^{2}}{\left(
I_{1}^{2}+I_{2}^{2}\right) ^{2}\left( \left( I_{1}-x_{p}\right)
^{2}+I_{2}^{2}\right) ^{2}}\right] \right\}
\end{multline}

This is the sought formula for the amplification of the images. Paczynski's
amplification multiplies the main brackets containing the sum of all
perturbations following the zero order solution represented by the unity.
Two kinds of perturbations can be recognized: the first is caused by the
previously found shift in the image positions $\mathbf{\Delta I}$; the
second is the consequence of the change of the function $\det J$ produced by
the presence of the planetary term in the lens equation. I have dropped the
modulus from the main brackets because its content is always positive since
the perturbations are smaller than unity (except for the zones where
perturbative method is no longer valid).

As usual, the validity of perturbation theory is limited to the regions
where perturbations are enough small to make sense. So it is necessary a
careful examination of the denominators of all perturbative terms. The shift
terms present the distance of the zeroth order image from the origin raised
to the sixth power. There's no problem for the principal image which is
always far beyond the Einstein ring, but this is not true for the secondary
image. However the ``failure'' rises in the limit of vanishing lensing where
the amplification of this image is so low to be totally masked by the
amplification of the principal image. When the amplification of the
secondary image begins to become important, the distance from the origin is
largely sufficient to eliminate all the problems and have fine
perturbations. The shift $\mathbf{\Delta I}$ becomes infinite when the
source passes through the origin; so the region very near the origin is the
first to avoid. The displacement $\mathbf{\Delta y}$ diverges when the
zeroth order image approaches the planet as could easily be foreseen for a
first order perturbation theory. As regards the terms coming from the
alteration of $\det J$, there's nothing new; the prescriptions are the same
as those from the other terms.

In sum we are allowed to use these amplification formulae for all source
positions being not too near the origin or generating images too close to
the planet. This hardly happens when the source is internal to the caustics.
We'll see that very reliable results can be obtained up to very short
distances from the centres of the caustics.

\subsection{Planetary image}

As previously announced, in this subsection I shall deal with the third
image. The presence of this image is absolutely tied to that of the planet.
Anyway, Paczynski's images can still provide a good starting point for our
analysis. In fact, if the planet is very far from the star, it too will
behave as a single lens. In this case, the planetary image is nothing but
the secondary Paczynski's image for a very low mass. In this limit, its
distance from the planet is of order $m_{2}$. So, in our perturbative
expansion, we are encouraged to search for images with distances from the
planet of order $m_{2}$. Let $\mathbf{I}^{p}$ be the position of the
planetary image. We have:
\begin{equation}
\mathbf{I}^{p}=\mathbf{x}_{p}+\mathbf{\Delta I}^{p}
\label{Image planetary expansion}
\end{equation}
with $\mathbf{\Delta I}^{p}$ of order $m_{2}$. Saving only the lowest
order, the lens application reads:
\begin{mathletters}
\begin{eqnarray}
y_{1} &=&x_{p}-\frac{m_{1}}{x_{p}}-\frac{m_{2}\Delta I_{1}^{p}}{\left(
\Delta I_{1}^{p}\right) ^{2}+\left( \Delta I_{2}^{p}\right) ^{2}} \\
y_{2} &=&-\frac{m_{2}\Delta I_{2}^{p}}{\left( \Delta I_{1}^{p}\right)
^{2}+\left( \Delta I_{2}^{p}\right) ^{2}}
\end{eqnarray}
\end{mathletters}

These equations can easily be solved. The solution is:
\begin{eqnarray}
I_{1}^{p} &=&x_{p}-\frac{m_{2}\left( y_{1}-y_{p}\right) }{\left(
y_{1}-y_{p}\right) ^{2}+y_{2}^{2}} \\
I_{1}^{p} &=&-\frac{m_{2}y_{2}}{\left( y_{1}-y_{p}\right) ^{2}+y_{2}^{2}}
\end{eqnarray}
where $y_{p}=x_{p}-\frac{m_{1}}{x_{p}}$\ is the zeroth order position of the
planetary caustic already rising in former discussions.

As ever the amplification is calculated by expanding (\ref{General
amplification}). The lowest order result is:
\begin{equation}
\mu _{I^{p}}=\frac{m_{2}^{2}}{\left[ \left( y_{1}-y_{p}\right) ^{2}+y_{2}^{2}%
\right] ^{2}}
\end{equation}

Notice how this formula is much more simple than other images amplification.

The denominators in these expressions vanish when $\mathbf{y}\rightarrow
\mathbf{y}_{p}$. Consequently the perturbative method fails when the
source is very close to the centre of the planetary caustic.

\subsection{Perturbative light curves in planetary microlensing}

Once we have found the amplification for each image, in order to obtain the
microlensing amplification map we must sum up the components coming from the
three images. However, we see that the contribution to the total
amplification of the planetary image is of the second order in $m_{2}$.
Since we are only considering first order corrections to Paczynski's curve,
this contribution is to be ignored. Therefore, from now on, we shall confine
ourselves to the principal and secondary images only.

One consideration is for the two hidden images coming out when the source
crosses a caustic. If the event regards the planetary caustic, the two
images can be found by carrying further the expansion (\ref{Image planetary
expansion}). The new images arise from higher order solutions and their
amplifications will also be of higher orders. So we don't worry about them.
On the contrary, if the source crosses the central caustic, the new images
appear near the star's Einstein ring, far from any possible starting point
for a perturbative expansion. As we are not taking them into account, we
cannot expect to obtain good results inside the central caustic. Anyway, central
caustic crossing events are very improbable, since the extension of this caustic
is $m_{2}^{2}/m_{1}^{2}$ times the star's Einstein ring.

Building light curves presents no difficulty. Chosen one source trajectory,
it suffices to parameterize $y_{1}$ and $y_{2}$ in the amplification map
properly. This is no longer a function of the radial coordinate because
there is no more rotational symmetry.

To account for the finite size of the source
a simple numerical integration of the perturbative amplification map on the source
area at each point of the trajectory can be performed. The curves thus obtained
can be compared to numerical ones given by ``inverse ray shooting'' algorithm.

All the results I show in this paper regard a system constituted by a star
with mass $m_{1}=1$ and a Jovian planet ($m_{2}=10^{-3}$). This choice has
been made in order to put in better evidence planetary perturbations and to
test the perturbative approach in the least favourable situation. Obviously
with Earth - like planets things can only go better.

\begin{figure*}
 \resizebox{12cm}{!}{\includegraphics{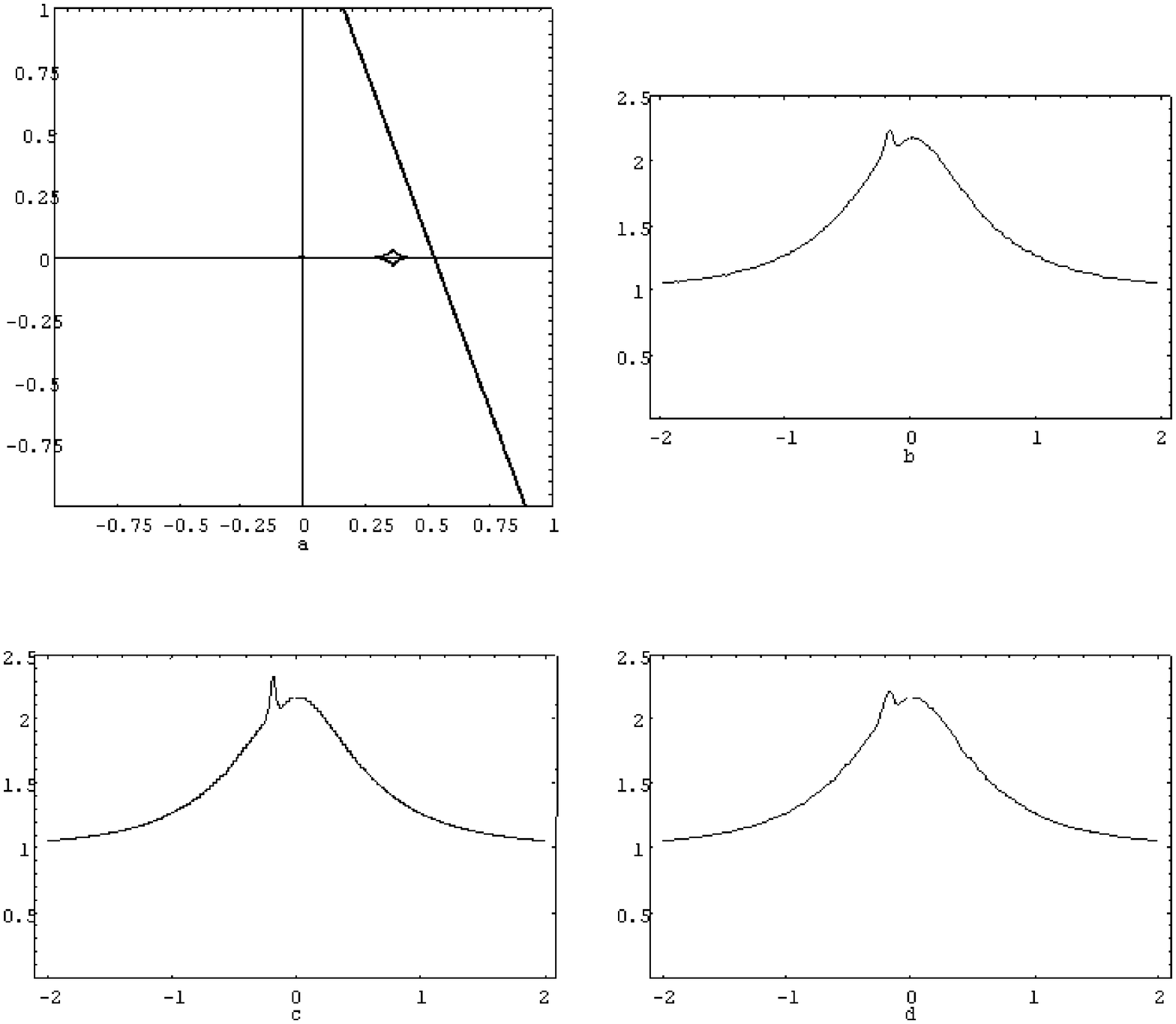}}
  \hfill
  \parbox[b]{55mm}{
   \caption{Besides a star with unitary mass
placed in the origin, here is a Jovian planet ($m_{2}=10^{-3}$) in $\protect%
\mathbf{x}_{p}=\left( 1.2;0\right) $. The trajectory of the source
shown in (a) has impact parameter 0.5. (b) is a numerical light
curve obtained by ``inverse ray shooting'' for a source 43 times
greater than the sun. (c) is the perturbative light curve for a
point source and (d) is the same curve after a numerical
integration over the source extension.}
  \label{Fig light
curve extern 0.5}}

\end{figure*}

\begin{figure*}
 \resizebox{12cm}{!}{\includegraphics{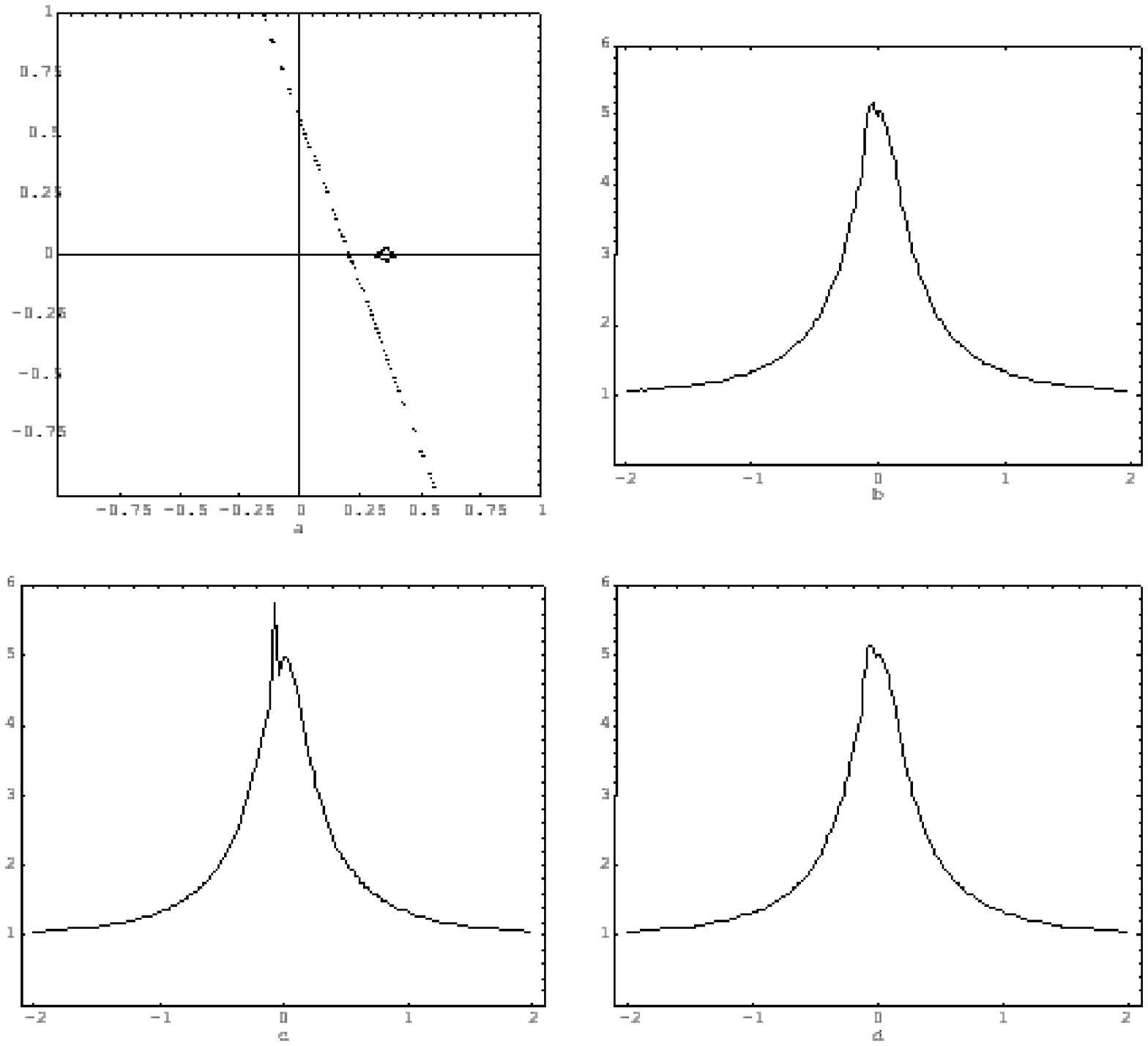}}
  \hfill
  \parbox[b]{55mm}{
   \caption{Same lens system and
same figure ordering as before. The impact parameter is 0.2.}
  \label{Fig light
curve extern 0.2}}

\end{figure*}

\begin{figure*}
 \resizebox{12cm}{!}{\includegraphics{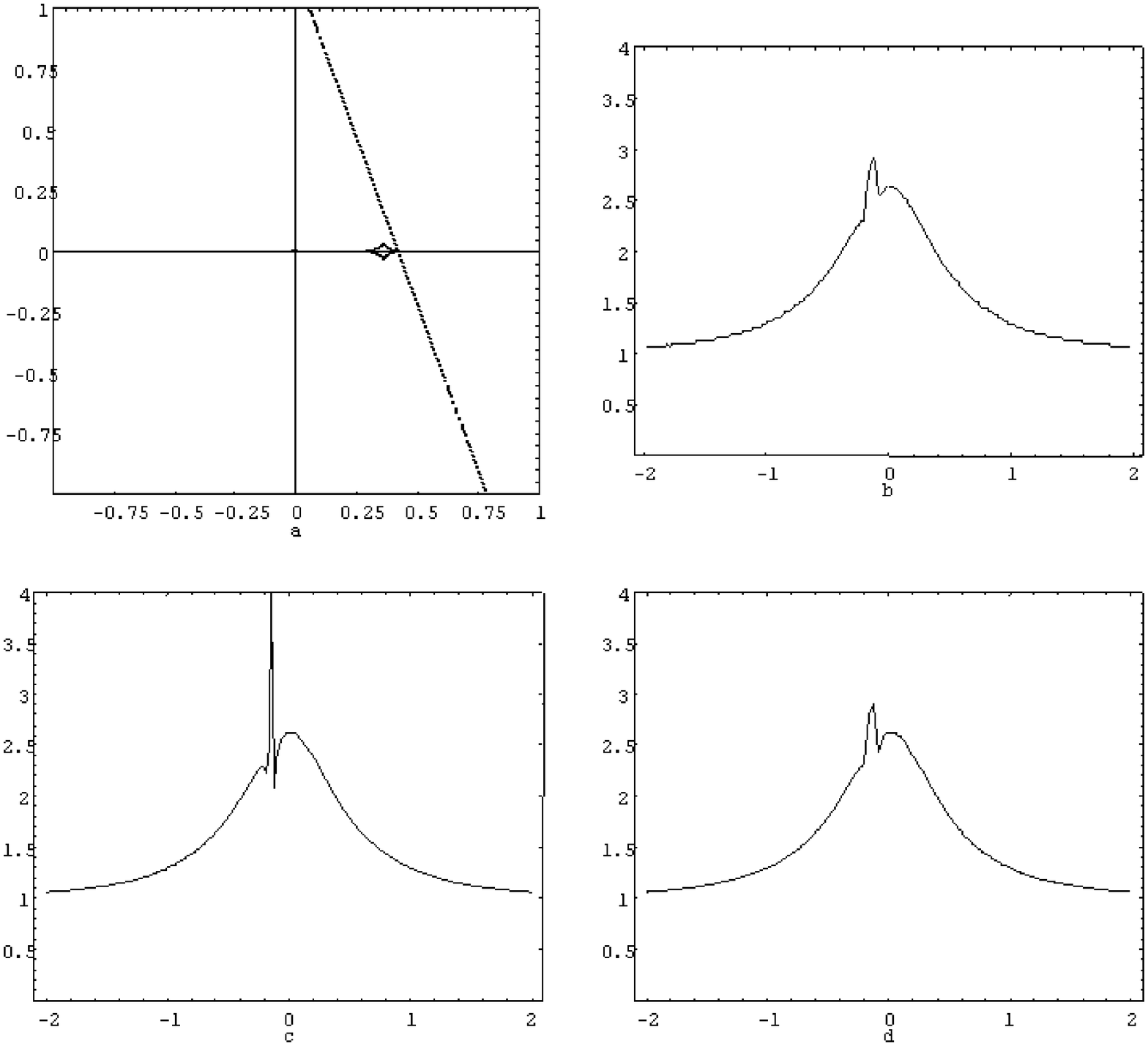}}
  \hfill
  \parbox[b]{55mm}{
   \caption{The lens system is the same
again but the impact parameter 0.4 makes the source cross the
planetary caaustic.}
  \label{Fig light
curve extern 0.4}}

\end{figure*}

\begin{figure*}
 \resizebox{12cm}{!}{\includegraphics{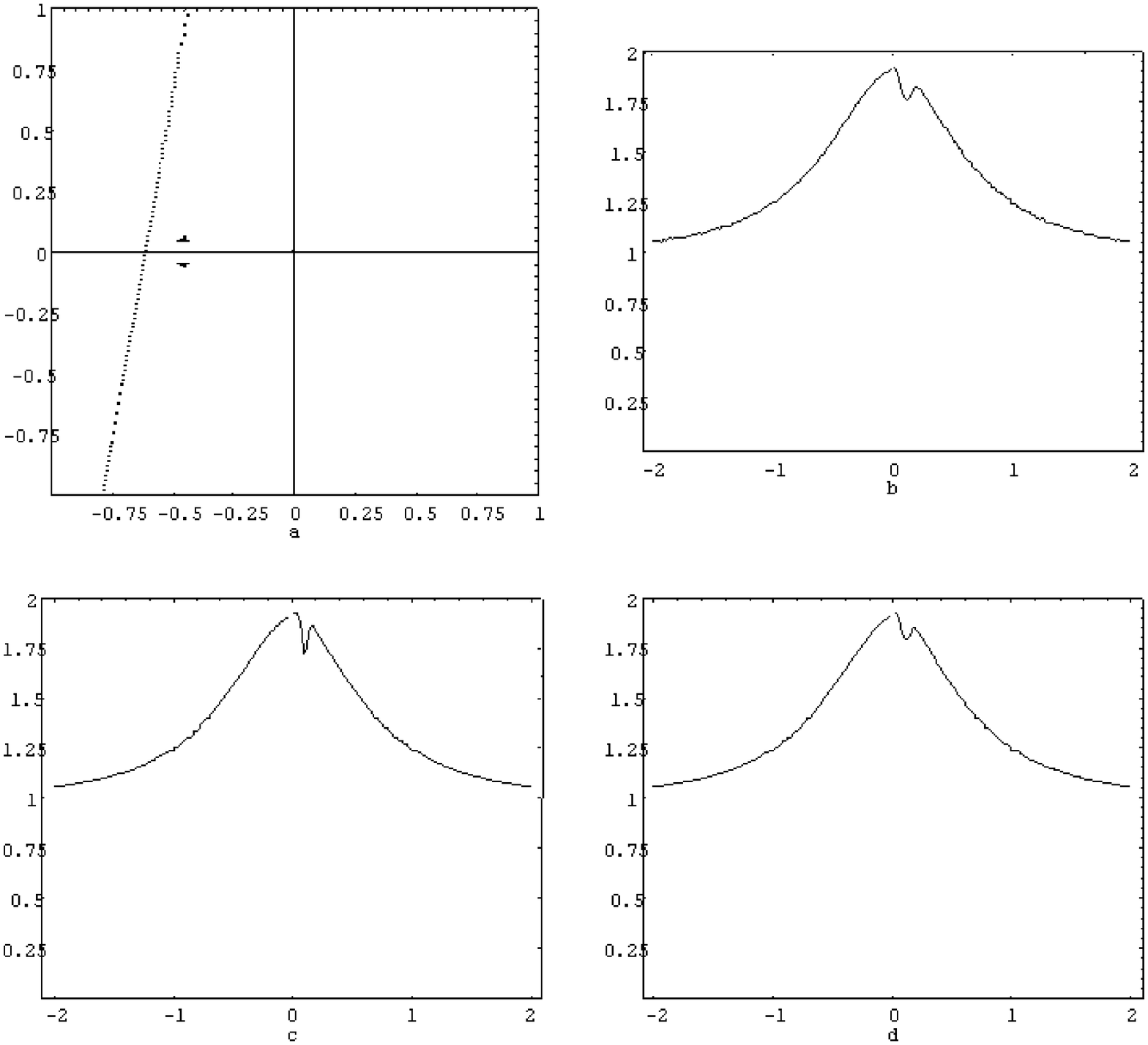}}
  \hfill
  \parbox[b]{55mm}{
   \caption{Now the planet is in $%
\protect\mathbf{x}_{p}=\left( 0.8,0\right) $. The three light
curves are
obtained in the same ways described in fig. \ref{Fig light curve extern 0.5}%
. The impact parameter is 0.6.}
  \label{Fig light curve internal 0.6}}

\end{figure*}

\begin{figure*}
 \resizebox{12cm}{!}{\includegraphics{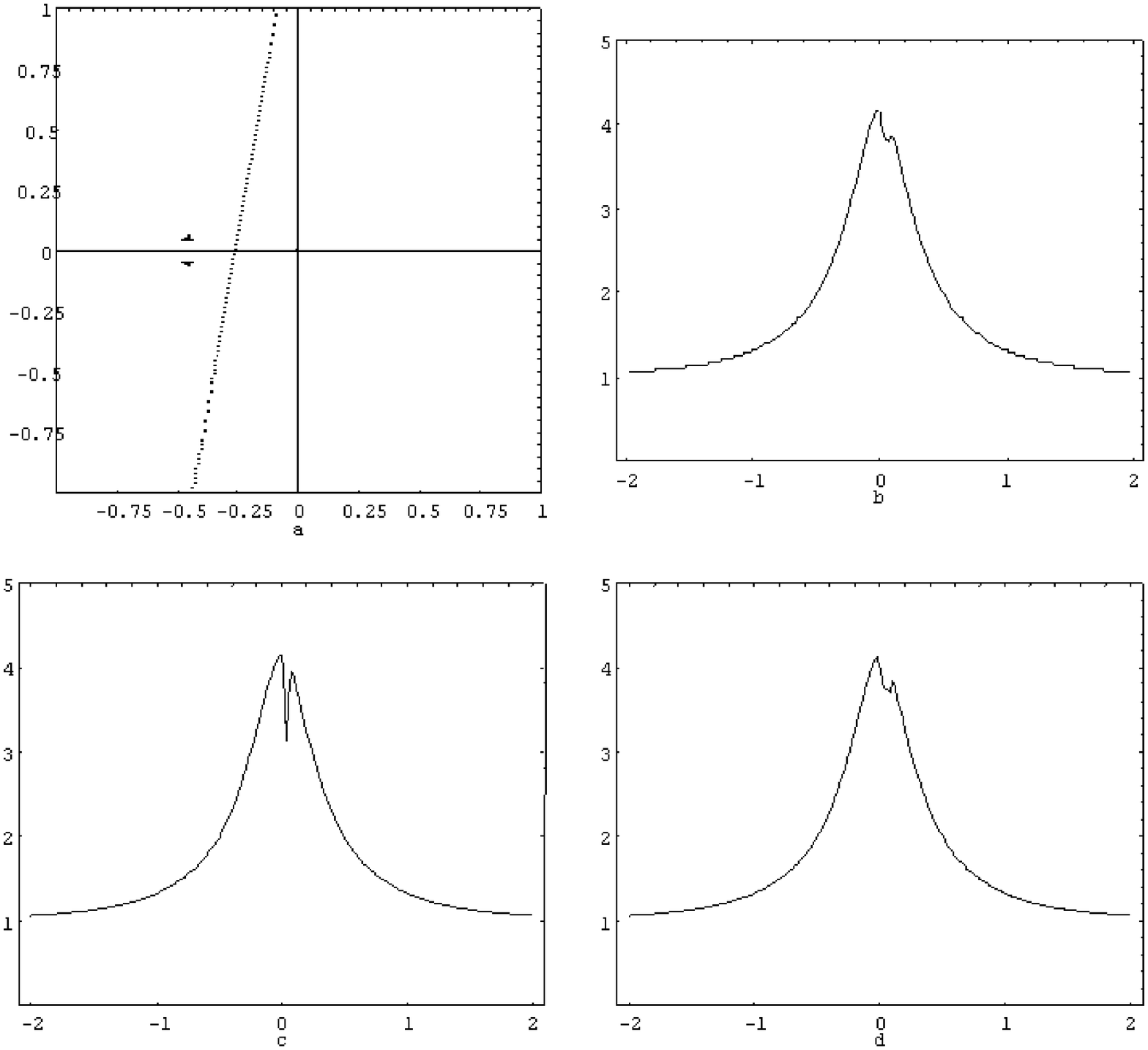}}
  \hfill
  \parbox[b]{55mm}{
   \caption{Same lens system as in
the previous figure. The impact parameter is 0.25.}
  \label{Fig light curve
internal 0.25}}

\end{figure*}

\begin{figure*}
 \resizebox{12cm}{!}{\includegraphics{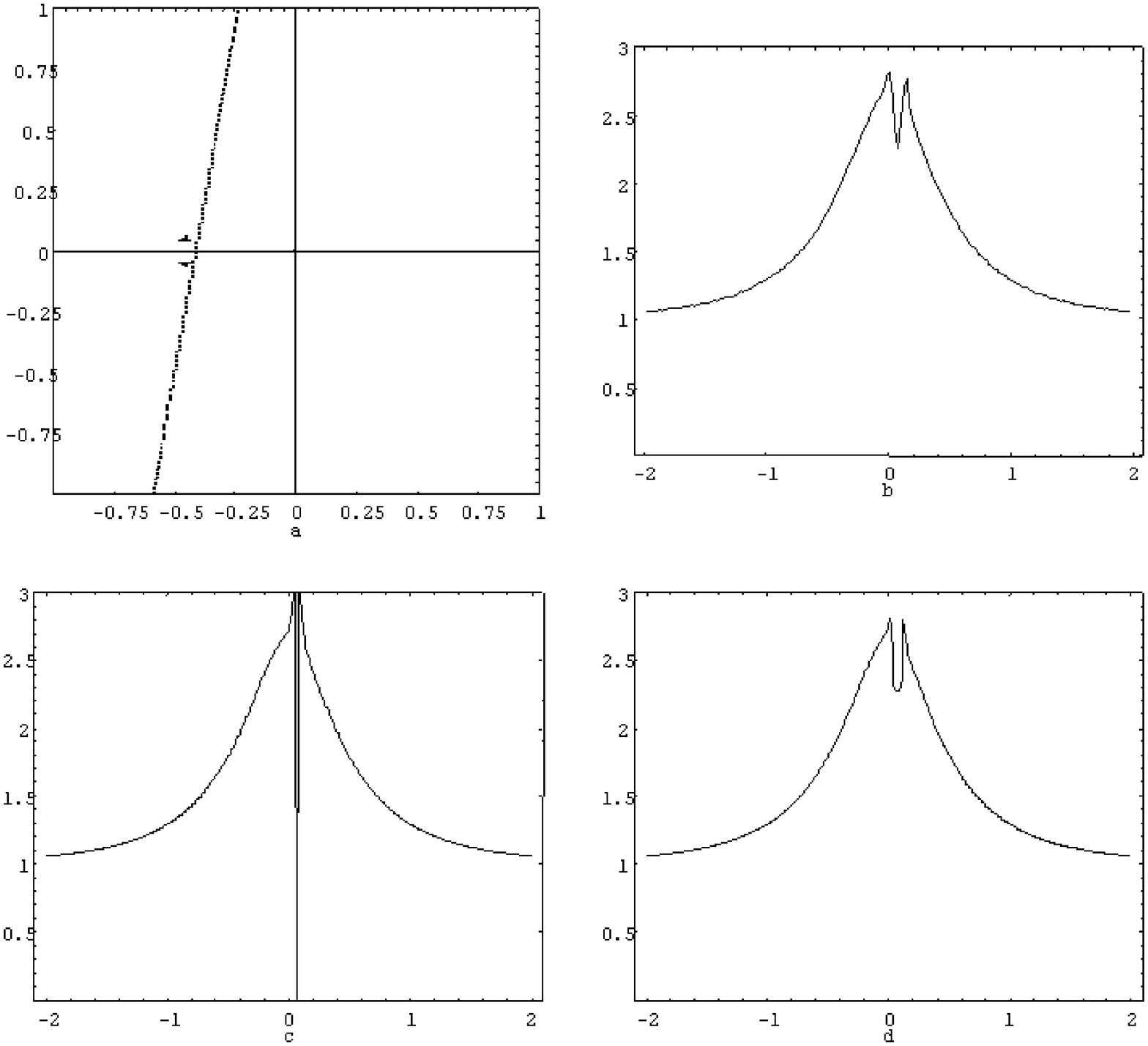}}
  \hfill
  \parbox[b]{55mm}{
   \caption{Caustic crossing with
impact parameter 0.4.}
  \label{Fig light curve internal 0.4}}

\end{figure*}

Let's start with an external planet. In fig. \ref{Fig light curve
extern 0.5} the planet is in $x_{p}=1.2$. The trajectory chosen
for this first test is shown in fig. \ref{Fig light curve extern
0.5}a and has impact parameter 0.5. The numerically attained light
curve is displayed in fig. \ref{Fig light curve extern 0.5}b. The
source used for this curve has radius 0.045.
In a standard observation towards the bulge of the galaxy ($D_{OL}\sim 8kpc$%
, $D_{OS}\sim 10kpc$), this value would correspond to a giant about 43 times
greater than the sun. Here the presence of the planet is responsible for the
little peak on the left of the maximum of the curve. Fig. \ref{Fig light
curve extern 0.5}c\ represents the perturbative light curve for a point
source moving along the same trajectory. If we perform the numerical
integration of the perturbative amplification map, as previously said, the
perturbative light curve \ref{Fig light curve extern 0.5}d becomes
indistinguishable from the numerical one.

This is a very encouraging result, so let's choose other trajectories to see
other tests. In fig. \ref{Fig light curve extern 0.2} the position of the
planet is the same but the trajectory passes between the planetary caustic
and the central caustic at a minimum distance of 0.2. The peak in the
numerical curve \ref{Fig light curve extern 0.2}b is very close to the
maximum. The point source perturbative curve \ref{Fig light curve extern 0.2}%
c presents a sharp peak which assumes the right proportions after the
integration in \ref{Fig light curve extern 0.2}d.

At this point, let's see what happens when the source crosses the planetary
caustic. In fig. \ref{Fig light curve extern 0.4}a, the impact parameter 0.4
allows the crossing. The peak in the numerical curve \ref{Fig light curve
extern 0.4}b becomes considerably high.\ In the point source perturbative
curve \ref{Fig light curve extern 0.4}c\ the peak is very sharp (it would
diverge at the centre of the caustic $\mathbf{y}_{p}$). However, the
integration over the source surface still succeeds in reporting this peak to
the right size and shape.

Now, let's consider an internal planet\ ($x_{p}=0.8$). The region between
the couple of planetary caustic is characteristic for its high
de-amplification. This produces negative peaks on light curves such as the
one shown in fig. \ref{Fig light curve internal 0.6}b corresponding to the
trajectory in fig. \ref{Fig light curve internal 0.6}a. It is interesting to
see that the perturbative method reproduces this situation with the same
great accuracy proved in the former situations. As ever, the point source
peak in \ref{Fig light curve internal 0.6}c is smoothed by finite source
effect in \ref{Fig light curve internal 0.6}d.

In\ fig. \ref{Fig light curve internal 0.25} the impact parameter is 0.25
and things go perfectly as previously.

Finally, let's consider caustic crossing in this case. Fig. \ref{Fig light
curve internal 0.4}a shows a trajectory very close to the planetary caustics.
The ``inverse ray shooting'' curve \ref{Fig light curve internal 0.4}b
presents a large de-amplification preceded and followed by little positive
peaks. The perturbative curve \ref{Fig light curve internal 0.4}c is
characterized by the same situation but the de-amplification is so high to
make the total amplification become (unphysically) negative. Now let's see
what happens with a finite source. Because of its extension, part of the
source hits the centre of the caustic $\mathbf{y}_{p}$ where the
perturbative amplification map diverges. This is a hard problem for the
numerical integration which becomes very unstable in this zone, so the
bottom of the de-amplification region of the light curve \ref{Fig light
curve internal 0.4}d cannot be taken as significant. However, we see that
things go fairly well even in this extreme situation.

\section{Conclusions}

\noindent The success of perturbative theory in planetary lensing cannot but
impress for the simplicity of the calculations involved and the surprising
accuracy of the results even in the hardest situations.

In the derivation of the caustics of a planetary system, by a simple idea
and very few passages the complete structure of these curves has been easily
obtained. The almost complete insensitivity of the perturbative approach to
the number of the planets allows complete descriptions of planetary systems
without any loss of generality. Also many important physical assertions can
be stated thanks to these results. The fact that the shape of the central
caustic is largely given by a linear superposition of the effects of \ the
single planets is indeed remarkable.

In planetary microlensing the results are even exalting. The perturbative
amplification map allows the construction of very fine light curves. In the
derivation of the amplification map I have dealt with only one planet for
the sake of simplicity. Yet the generalization to an arbitrary number of
planets is immediate because in the first order domain a superposition
principle is here valid as well. For point sources, light curves can be
attained in a completely analytical way, while for finite sources I have
resorted to numerical integration until now. Work to englobe finite source
effect in the analytical description is in progress. When these curves are
available, the extraction of parameters of planetary systems from
microlensing light curves will start on more solid analytical bases. Also it could be
possible to use the analytical expressions in experimental fits, though the
large number of parameters would greatly affect the uncertainties in their
determination.

\begin{acknowledgements}
I would like to thank Salvatore Capozziello, Gaetano Lambiase and Gaetano
Scarpetta for their valuable suggestions and interesting discussions on
this matter. Also I greatly wish to acknowledge Giovanni Covone for guiding
me in my initiation to planetary microlensing.
\end{acknowledgements}

\end{document}